\documentclass[preprint,showpacs, aps, prb, amsmath,amssymb]{revtex4}

\usepackage[dvips]{graphicx}
\usepackage{dcolumn}
\usepackage{bm}

\begin{document}

\preprint{}

\title{Optical probe of electrostatic doping in an $n$-type Mott insulator}

\author{M.~Nakamura}
\email{m-nakamura@aist.go.jp}
\affiliation{%
Correlated Electron Research Center (CERC), National Institute of
Advanced Industrial Science and Technology (AIST), Tsukuba 305-8562, Japan
}%
\affiliation{%
Core Research for Evolutional Science and Technology (CREST), Japan
Science and Technology Agency (JST), Kawaguchi 332-0012, Japan
}%
\author{A.~Sawa}\affiliation{%
Correlated Electron Research Center (CERC), National Institute of
Advanced Industrial Science and Technology (AIST), Tsukuba 305-8562, Japan
}%
\affiliation{%
Core Research for Evolutional Science and Technology (CREST), Japan
Science and Technology Agency (JST), Kawaguchi 332-0012, Japan
}%
\author{H.~Sato}
\affiliation{%
Correlated Electron Research Center (CERC), National Institute of
Advanced Industrial Science and Technology (AIST), Tsukuba 305-8562, Japan
}%
\affiliation{%
Core Research for Evolutional Science and Technology (CREST), Japan
Science and Technology Agency (JST), Kawaguchi 332-0012, Japan
}%
\author{H.~Akoh}
\affiliation{%
Correlated Electron Research Center (CERC), National Institute of
Advanced Industrial Science and Technology (AIST), Tsukuba 305-8562, Japan
}%
\affiliation{%
Core Research for Evolutional Science and Technology (CREST), Japan
Science and Technology Agency (JST), Kawaguchi 332-0012, Japan
}%
\author{M.~Kawasaki}\affiliation{%
Correlated Electron Research Center (CERC), National Institute of
Advanced Industrial Science and Technology (AIST), Tsukuba 305-8562, Japan
}%
\affiliation{%
Institute for Materials Research, Tohoku University, Sendai 980-8577, Japan
}%
\author{Y.~Tokura}\affiliation{%
Correlated Electron Research Center (CERC), National Institute of
Advanced Industrial Science and Technology (AIST), Tsukuba 305-8562, Japan
}%
\affiliation{%
Department of Applied Physics, University of Tokyo, Tokyo 113-8656, Japan
}%
\date{\today}

\begin{abstract}
Electrostatic doping into an $n$-type Mott insulator Sm$_{2}$CuO$_{4}$
 has been successfully achieved with use of the heterojunction with an
 $n$-type band semiconductor Nb-doped SrTiO$_{3}$. The junction exhibits
 rectifying current-voltage characteristics due to the interface band
 discontinuity and the formation of depleted region. The application of
 reverse bias electric field on this junction enables the field-effect
 electron doping (presumably up to 6\% per Cu atom) to the Mott
 insulator. The electro-modulation absorption spectroscopy could clearly
 show a large modification of the Mott-gap state accompanying the
 spectral weight transfer to the lower-energy region, reminiscent of
 formation of a metallic state.
\end{abstract}

\maketitle
\section{Introduction}
The electronic structure of parent Mott insulators
and its dramatic modification with carrier doping, especially at the
phase boundary, is of crucial importance for the fundamental
understanding of the phase transition in strongly correlated electron
materials~\cite{imada}. 
The electrostatic carrier doping is an attractive methods for such
a study because it enables us a critical control of the carrier density
without introducing randomness into the system.
Usually, the electrostatic doping has been
performed with use of the field-effect transistors (FETs)
with metal-insulator-semiconductor (MIS) structures illustrated in
Fig.~1(a)(b)~\cite{newns,ahn}. 
In this structure, it is essential to prepare a high quality insulating film
which should have large breakdown field and small leakage current to 
induce a change of carrier density (band filling)
enough for the phase transition~\cite{mis}. However, a number of 
previous works have proved it technically difficult, while 
the problem of the leakage current could be partly overcome with using
a ferroelectric compound as a gate insulator~\cite{ahn,ahn2}. 

To overcome the difficulty, we adopt another scheme of field-effect
device in 
this study, that is an isotype heterojunction comprised of a Mott
insulator and a semiconductor (Fig.~1 (c)).
A typical energy-band diagram of an $n$(narrower-gap
semiconductor)-$N$(wider-gap semiconductor) isotype heterojunction 
as InGaAs/GaAs is shown in Fig.~1(d)~\cite{capasso}.
Due to the difference in the chemical potential between these two
contacting materials, the conduction-band discontinuity and band bending
emerge at the interface. The band bending yields the charge depleted and
accumulated regions at the interface.
The self-formed depleted region works as a capacitor in such a
junction, ensuring the high experimental 
reproducibility coupled with its simple device structure.
The induced interface carrier density can be estimated
from capacitance-voltage measurements.
By replacing the narrow-gap semiconductor with an $n$-type Mott insulator,
it is expected to inject charge carriers into the Mott insulator with the
field-effect. Here, ``$n$-type Mott insulator'' means the Mott insulator
in which the band filling can be increased to produce the conducting
state, like $T'$-phase $R_{2}$CuO$_{4}$ ($R=$Pr, Nd, and Sm) known as the
parent insulators of electron-doped high-$T_{c}$ superconductors.

\begin{figure}[htbp]
\centering
 \includegraphics[width=7cm]{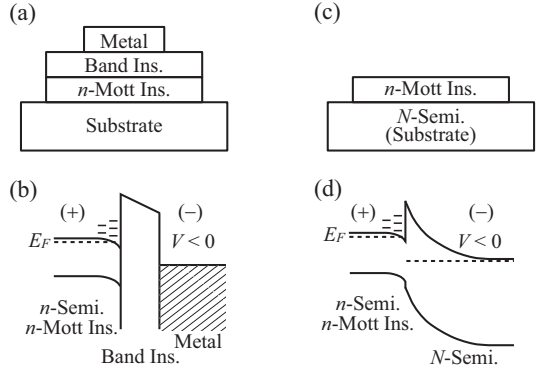}
 \caption{ Schematics of the device structures for the field-effect
 carrier doping to an $n$-type Mott insulator with a MIS structure (a) and
 with an isotype heterojunction (c). The abbreviations, Ins. and Semi., in
 the figure denote the insulator and the semiconductor, respectively. 
 Typical band diagrams of a junction with the MIS structure (b)
 and that of the $n$-$N$ isotype heterojunction (d). The dashed lines
 represent Fermi-energy.}
\end{figure}

In this Letter, we report on the electrostatic carrier
doping to a Mott insulator and its optical detection in the
heterojuncion comprised of Nb-doped 
SrTiO$_{3}$ (Nb:STO) and Sm$_{2}$CuO$_{4}$ (SCO).
The crystal structure of SCO is the so-called $T'$-phase composed of the
alternate stack of fluorite-type Sm$_{2}$O$_{2}$ layers and CuO$_{2}$
sheets with the lattice 
constant of $a=3.914$~\AA~and $c=11.972$~\AA~\cite{yang}. Since the
$a$-axis lattice constant is close to that of STO (3.905~\AA), a
pseudomorphic interface is expected to reduce possible 
defects originating from the lattice mismatch.
Nb:STO is an $n$-type semiconductor with a band-gap of 3.2
eV. The energy-band diagram of the SCO/Nb:STO
heterojunction was evaluated from current($I$)-voltage($V$) and
capacitance($C$)-voltage($V$) 
measurements. As a method for the observation of the electrostatic carrier
doping, we adopted the electro-absorption (EA) spectroscopy. This technique
extracts only the modulated component in the absorption through the 
application of an electric field~\cite{cardona}. In the heterojunction,
the field-induced absorption change would be limited in the vicinity
of the interface. Thus, we expected that the EA spectroscopy serves as
a sensitive probe of the injected carriers to the Mott insulator at
the heterointerface. We stress here that the spectroscopic observation
allows us to extract information on the variation of the electronic
structure of the Mott insulator.

\section{Experiments}
A 30 nm-thick SCO film was grown on a (100) Nb:STO
substrate held at 780$^{\circ}$C in an oxygen pressure of 300 mTorr by
pulsed-laser deposition, and the films was subsequently annealed to romove
excess apical oxygens~\cite{sawa}.
The sample should be semi-transparent so as
to measure optical spectra in transmission geometry. Therefore,
we employed a Nb 0.01 wt\% doped STO (carrier density of
1$\times10^{18}$ cm$^{-3}$) as the
substrate, which is almost transparent in the energy region of
the visible light, although it has a metallic conductivity.
X-ray diffraction measurements  show that the film has a single-phase
$c$-axis oriented pseudomorphic structure ($c=11.99$~\AA).
The lattice constants of the film are close to those of the
bulk~\cite{yang}, signifying that the film is in $T'$-phase as expected.

To obtain an ohmic contact to the SCO and to apply an uniform electric field
throughout the measurement area, semi-transparent Au film with a
thickness of 6 nm was \textit{ex situ} evaporated on top of the SCO film. 
The Au/SCO/Nb:STO junction was then patterned into a
mesa-structure (the inset of Fig. 3) by photo-lithography,
Ar ion milling, sputter coating of SiO$_{2}$, 
and lift-off procedures. Finally, a 300 nm-thick Au film was deposited and
patterned so as to work as the contact to the thinner Au film as well as
lead wires. 
The size of the junction area is $200\times 200~\mu\mathrm{m}^{2}$
on which a monochromatic incident light was focused by an objective
lens for spectroscopy. 
The base electrode was subsequently provided by depositing Al film which can
form an ohmic contact to Nb:STO.

\section{Results and Discussion}
The $I$-$V$ characteristics of an SCO/Nb:STO junction
is shown in Fig.~2 (on the left ordinate).
The junction shows a rectification property without
a breakdown up to a reverse bias of $-10$~V.
The rectification property indicates the existence of the band discontinuity
and depleted region. 
Possible energy-band diagrams of SCO/Nb:STO junction is shown in the
inset of Fig.~2 by analogy with a semiconductor $n$-$N$ isotype
heterojunction. 
The depletion length in Nb:STO reaches several hundreds nanometers in our
case due to the large permittivity ($\varepsilon_{\mathrm{Nb:STO}}\sim300$)
and the small doping concentration. 
This rather simple rigid-band picture is strictly not the case for strongly
correlated electron systems including Mott insulators in which
the band reconstruction occurs with the carrier doping or with change of
band filling. We believe, however, the conduction-band
discontinuity and the formation of the depleted and accumulated regions
are crucial to realize such rectification property in the SCO/Nb:STO
heterojunction. 

To obtain more quantitative information about the interface
band discontinuity, we measured capacitance of the same junction.
For an ideal $n$-$N$ heterojunction, $C$-$V$ characteristics is given
by~\cite{capasso} 
\begin{equation}
 \frac{1}{C^{2}}=\frac{2kT}{q\varepsilon
  _{2}N_{2}}\frac{q(V_{D2}-V_{2})/kT-1}{1-(kT/q)\left[(V_{D}-V)-(V_{D1}-V_{1})(1-\varepsilon_{1}N_{1}/\varepsilon_{2}N_{2})\right]^{-1}}, 
\end{equation}
where $\varepsilon$, $N$, and $V_{D}$ denote permittivity, carrier density,
and diffusion potential, respectively, and the subscripts 1, 2 denote
SCO and Nb:STO, respectively.
If $V\gg kT/q$ and 
$\varepsilon_{1}N_{1}/\varepsilon_{2}N_{2}\gtrsim1$, Eq. (1) can be 
approximated by~\cite{forrest}
\begin{equation}
 \frac{1}{C^{2}}=\frac{2}{q\varepsilon_{2}N_{2}}\left(V_{D}-V\right).
\end{equation}
Equation~(2) is the same formula as that for a 
Schottky junction, and we can consider the band-offset on the analogy of
that for a Schottky barrier. 
In Fig.~2 (on the right ordinate), we show a $1/C^{2}$-$V$ curve measured at a
frequency of 1 MHz.  
It is readily apparent that the plot of $1/C^{2}$ versus $V$ shows a linear
relationship, indicating a small contribution from the trapped charges
on the interface defects. 
$V_{D}$ is found to be 1.3 V from the voltage-axis
intercept.
It is difficult to deduce an accurate value of the conduction-band
offset ($\Delta E_{c}$) because of a little 
information on the position of the Fermi energy relative to the
conduction-band in SCO ($\delta_{1}$).
According to the result of x-ray photoemission spectroscopy for
Nd$_{1.85}$Ce$_{0.15}$CuO$_{4}$~\cite{taguchi}, 
$\delta_{1}$ is so small as to assume that $\Delta{E_{c}}\approx V_{D}$.

\begin{figure}[tbp]
\centering
 \includegraphics[width=8cm,clip]{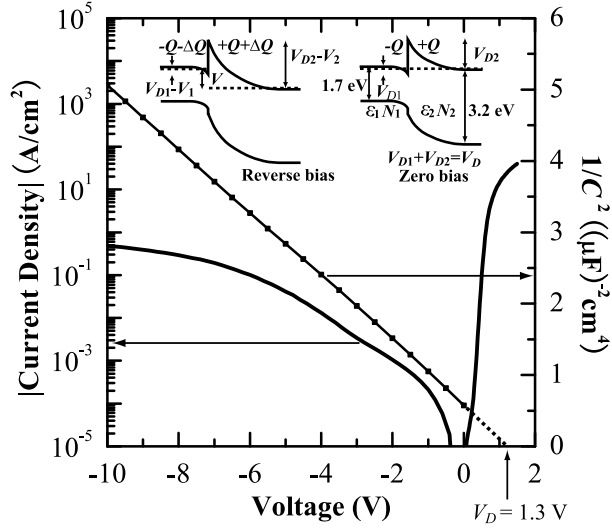}
\caption{Current-voltage and capacitance-voltage
 characteristics of a heterojunction comprised of Sm$_{2}$CuO$_{4}$
 (SCO) and 0.01wt\% Nb-doped SrTiO$_{3}$ (Nb:STO).
 The inset shows possible energy-band diagrams of the SCO/Nb:STO
 junction at zero-bias (right) and under reverse bias (left) states.
 $\varepsilon$, $N$, $V_{D}$ denote permittivity, carrier density, and
 diffusion potential, respectively, and the subscripts 1 and 2 denote SCO
 and Nb:STO, respectively.}
\end{figure}

As illustrated in the inset of Fig.~2, the application of reverse bias
should induce the accumulation of additional electrons in
SCO at the interface in compensation for further exposed positive
space-charge of ionized donors in depleted Nb:STO.
The EA measurements were
performed on the device with the configuration shown in the
inset of Fig.~3. 
Rectangle wave electric field (570 Hz) with various reverse biases was
applied to the specimen with a pulse generator. 
The semi-transparent area of the device was illuminated with a
monochromatic light from a halogen or a xenon lamp,
and the transmitted light was detected with a Si photodiode
(1.1-3.2 eV) or an InAs photovoltaic detector (0.5-1.1 eV).
The modulated component of the transmitted light ($\Delta T$)
synchronized with the electric field was measured with a lock-in
amplifier while monitoring the constant component ($T$).
The relation between  $\Delta T/T$ and the change in the absorption
induced by the electric field is approximated by
$\Delta T/T=-\Delta(\alpha d)$,
if $|\Delta T/T|\ll 1$, where $\alpha$ is the absorption coefficient and 
$d$ is the thickness over which the spectral change occurs~\cite{cardona}.

The optical absorption spectrum of a thin film of SCO is shown in Fig.~3
(on the right ordinate).
The spectrum has a sharp absorption peak at 1.7 eV and a
broad side peak at 2.2 eV, similarly to the bulk feature~\cite{arima}. 
The main absorption peak corresponds to the charge-transfer (CT)
transition from the O~2$p$ band to the Cu~3$d_{x^{2}-y^{2}}$ upper
Hubbard band (UHB)~\cite{tokura2}, and
its fairly sharp profile indicates the exitonic character of the
transition. Within the CT gap, optical absorption is negligibly small.
The EA spectra of the SCO/Nb:STO junction are shown in Fig.~3 (on the
left ordinate). 
A very systematic enhancement of EA spectra is observed
with increasing the amplitude of voltage modulation.
At a first glance, one can notice that the spectra exhibit a sign reversal
near the absorption-edge. 
The absorption increases ($\Delta \alpha >0$)
below the absorption-edge, whereas the CT absorption decreases ($\Delta
\alpha <0$) above that. 
To consider the origin of this feature,
let us recall the effect of the carrier doping on the optical spectrum
for under-doped $n$-type cuprates.
The previous studies with use of chemically doped bulk samples have
revealed that the electron doping gives rise to the emergence 
of a new state within the CT gap and the suppression of the spectral
weight in the higher energy region~\cite{cooper,uchida,arima2}. 
For comparison, we also fabricated a 2\% Ce doped
(Sm$_{1.98}$Ce$_{0.02}$CuO$_{4}$) thin film and measured the
absorption spectrum as shown in Fig.~4(a).  
The close agreement of the EA spectra with the
doping induced spectral change clearly indicates that 
the observed sign reversal in the EA spectra is a
consequence of the field-effect electron doping in SCO.
One might dispute that the EA spectra is caused by a low-energy shift of
the absorption-edge, that would be the Franz-Keldysh effect as observed
for semiconductors or band insulators~\cite{cardona}. If this were the
case, the EA 
spectra would show substantial signals only near the absorption-edge and
the oscillating structure of the EA spectra would show a large field
dependence.
However, we observed fairly large signals far below the absorption-edge,
where no optical transition is observed in the linear absorption spectrum.
Therefore, the influence of the red-shift of the absorption-edge is
not a major source, if any. 
The existence of the isosbestic (equal-absorption) point at 1.47 eV is
also an evidence 
of the transfer of the spectral weight from the CT absorption in a
higher energy region to the in-gap absorption in a lower energy region.

\begin{figure}[tbp]
\centering
 \includegraphics[width=8cm,clip]{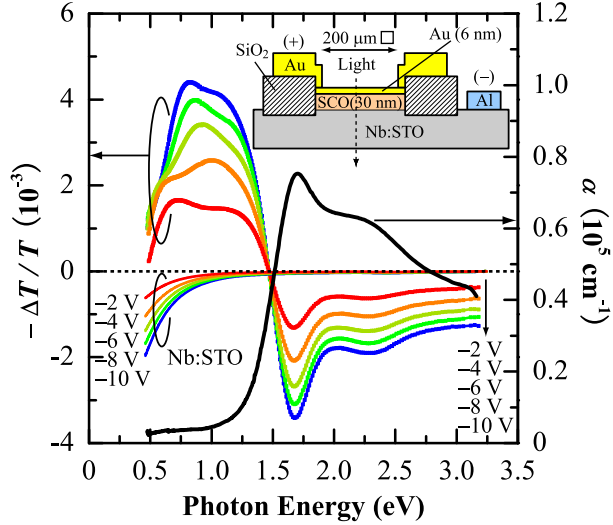}
\caption{Ordinary optical absorption spectrum of 
 an SCO film (on the right ordinate), and electro-absorption
 (EA) spectra of the 
 SCO/Nb:STO heterojunction 
 measured under various reverse bias voltages (on the left
 ordinate). Estimated contribution from the depleted region of Nb:STO at
 each bias voltage 
 is also shown with the lines denoted as Nb:STO. The inset depicts the
 cross-sectional view of the device structure used in $I$-$V$, $C$-$V$,
 and EA measurements.}  
\end{figure}

Next, we discuss the spectral feature below the absorption-edge.
As shown in the inset of Fig.~2, the application of reverse bias results
not only in the accumulation of charges in SCO but also in the
widening of the depleted region in Nb:STO. Therefore, the
EA spectra should include the contribution from
bleaching of free carrier absorption in Nb:STO.
We estimated the Nb:STO contribution to the EA spectra from the quantity,
$(\alpha_{\mathrm{STO}}-\alpha_{\mathrm{Nb:STO}})\Delta W$. Here,
$\alpha_{\mathrm{STO}}$ and $\alpha_{\mathrm{Nb:STO}}$ are absorption
coefficients of non-doped STO and Nb:STO, respectively, and $\Delta W$ is
the expanded depletion length by the applied electric field as deduced
from the $C$-$V$ 
measurement. As shown in Fig.~3, the contribution from Nb:STO is appreciable
only below the mid-infrared region ($\hbar\omega<1~\mathrm{eV}$), and
the influence can be understood as a reduction of Drude-like free carrier
absorption. Therefore, the
spectral drop toward zero as observed in the lower energy region of the
EA spectra may partly come from this bleaching effect of Nb:STO.
By subtracting the Nb:STO contribution from the EA
spectra, the field-effect contribution in SCO can be derived as shown in
Fig.~4(b) (only the case of $V=-2$~V is shown). 
After the subtraction, the feature of the spectral drop at lower energy
still exists. We speculate that the drop may indicate
the formation of the pseudogap originating from antiferromagnetic spin
correlation as observed in the bulk under-doped
Nd$_{2-x}$Ce$_{x}$CuO$_{4}$~\cite{onose}.

The carrier density injected by the
electric field can be obtained from the equation, $\Delta
Q=N_{2}\Delta W$. The estimated $\Delta Q$ is
$2.1~\mu\mathrm{C/cm^{2}}$ (0.020 electrons per unit cell area) at 
$V=-2$~V and $6.7~\mu\mathrm{C/cm^{2}}$ (0.063 electrons per unit cell
area) at $V=-10$~V. These values are comparable to the aforementioned
carrier density attainable in the MIS structure.
It is worth comparing the absorption changes induced by
electrostatic-doping ($\Delta\alpha_{\mathrm{E}}$) and chemical-doping
($\Delta\alpha_{\mathrm{\mathrm{C}}}$), respectively, at the same 
doping level of 2\%. 
$\Delta\alpha_{\mathrm{E}}$ is deduced from the EA spectrum obtained at
$V=-2$~V corrected by the contribution from Nb:STO, and assuming that the
doping was effective only 
for a single CuO$_{2}$ plane, $i.e.$, $d=c/2=0.6$~nm.
$\Delta\alpha_{\mathrm{C}}$ was calculated from the 
difference of absorption spectra between Ce 2\% doped SCO and undoped SCO. 
As shown in Fig.~4(b), the spectrum of
$\Delta\alpha_{\mathrm{E}}$ has similar profile and
magnitude with those of $\Delta\alpha_{\mathrm{C}}$.
The result ensures that the injected carriers are confined within
quite narrow region, possibly in a single CuO$_{2}$ plane adjacent to the
interface. This interpretation is reasonable if one
considers that the typical screening length in a Mott insulator is a few
nanometers~\cite{ohtomo, okamoto} and electron doped cuprates show quite a
larger effective mass along the $c$-axis direction than that within
$ab$-plane~\cite{beom}. 

\begin{figure}[tbp]
\centering
 \includegraphics[width=7cm,clip]{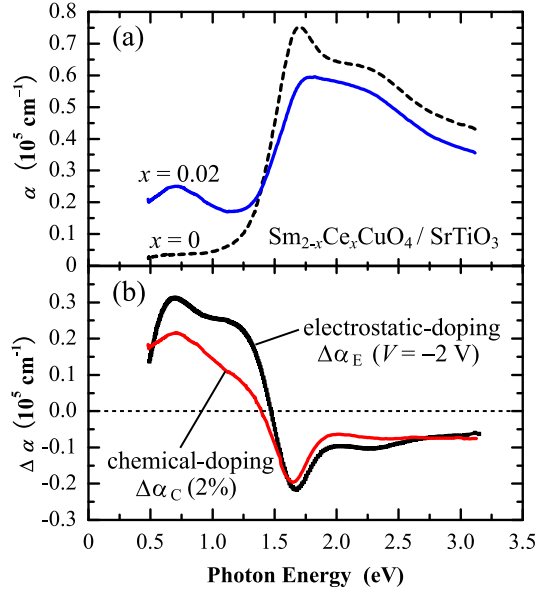}
\caption{(a) Ordinary optical absorption spectra of
 Sm$_{2-x}$Ce$_{x}$CuO$_{4}$ 
 ($x=$0 and 0.02) thin films grown on STO substrates. (b) Electrostatic-
 and chemical-doping induced changes of absorption spectra.
 The electrostatic-doping induced change ($\Delta\alpha_{\mathrm{E}}$)
 is deduced 
 from the EA spectrum measured at a bias voltage of
 $-2$~V and corrected by the contribution from Nb:STO. The doping region
 is assumed to be confirmed within the top-most layer of SCO (see text).
 The chemical-doping induced change ($\Delta\alpha_{\mathrm{C}}$) is
 obtained from the difference of absorption spectra shown in (a).}
\end{figure}

\section{Conclusions}
We have detected the electric field induced change in
electronic structure of a 
Mott insulator by an optical probe for a heterojunction
comprised of Sm$_{2}$CuO$_{4}$ and Nb-doped SrTiO$_{3}$, namely, an
$n$-type Mott insulator and an $n$-type 
semiconductor. This junction exhibits a rectifying current-voltage
characteristics due to the conduction-band discontinuity of about 1.3 eV as
revealed by the capacitance-voltage measurement.
The electro-absorption spectroscopy under reverse bias unraveled
the collapse of charge-transfer gap and the emergence of in-gap states via
electric-field doping of electrons in the top-most layer of
Sm$_{2}$CuO$_{4}$. 
A concept of $n$(Mott insulator)-$N$(semiconductor) or $p$(Mott
insulator)-$P$(semiconductor) isotype heterojunction is applicable as an
effective carrier doping method to a wide range of Mott insulators.
The interface-sensitive electro-modulation spectroscopy may
quantitatively elucidate the electronic state of the Mott insulator at
the heterointerface while continuously changing the doping level or band
filling. 

\begin{acknowledgments}
 We are grateful to H. Matsuzaki, Y. Okimoto, and T. Ogasawara for
 technical assistance, and to Y. Ogimoto, T. Hasegawa, and H. Okamoto for
 helpful discussions. 
\end{acknowledgments}


\begin{thebibliography}{99}
\bibitem{mis} For example, when  Al$_{2}$O$_{3}$ is used as the gate
	 insulator, the estimated carrier density is 7.5
	 $\mu\mathrm{C/cm}^{2}$, corresponding to 0.07 electrons per the
	 area of 1 unit cell squared of a perovskite oxide, under an
	 electric field of 10 MV/cm.
\end{thebibliography}
\end{document}